\documentclass[11pt]{article}
\usepackage{hyperref}
\pdfoutput=1

\begin{document}

\title{Instability of liquid jet\\ penetrated into stream in channel}

\author{Naoto Oka$^{1}$, Ichiro Ueno$^{2}$ \\
\\\vspace{0pt} $^{1}$Graduate School, Tokyo University of Science, \\\vspace{6pt} 2641 Yamazaki, Noda, Chiba 278-8510, JAPAN
\\\vspace{6pt} $^{2}$Tokyo University of Science, Noda, JAPAN}

\maketitle


\begin{abstract}
Penetration process and an instability on a liquid jet impinging to a stream of the same fluid in a channel is focused. The jet penetrated
into the stream is wrapped by entrained air, and coalesces with the stream when the air sheath around the jet collapses. We introduce
instability arisen on the jet and the vigorous effect of the entrained-air sheath on the dynamic behavior of the jet in this fluid dynamics video.

\end{abstract}


\section{Introduction}




 In the present fluid dynamics \href{http://ecommons.library.cornell.edu/bitstream/1813/14090/2/APS-fluid-of-motion-oka.mpg}{video}, 
 we focus on instability of the penetrated jet of 100-cSt
silicone oil to the same liquid flowing in the channel. The penetrated
jet exhibits a Rayleigh-Plateau-like instability to break up into
droplets. We are interested in a unique behavior of the jet affected by
the entrained air; once the broken tip of the jet is completely capped
by chance by the air film around the jet, the disturbance is
unexpectedly destabilized to form the stable jet penetrating further
without breakdown. This instability might be essential to realize the
jet bouncing off from the fluid flowing in the channel (1).
\\REFERENCES
\\(1) Thrasher, M. et al., Phys. Rev. E 76,056319, 2007.
%
\end{document}